\documentclass[twocolumn,draft,prd,nofootinbib]{revtex4} 
\usepackage{epsf}% Include figure files 
\usepackage{dcolumn}% Align table columns on decimal point 

\newcommand{\be}{\begin{equation}} 
\newcommand{\ee}{\end{equation}} 
\newcommand{\Deln}{\ensuremath{\Delta N_\nu\;}} 
\newcommand{\epm}{\ensuremath{e^{\pm}\;}} 
 
\def\ie{{\it i.e.},~} 
\def\eg{{\it e.g.},~} 
\def\etal{{\it et al.}~} 
\def\4he{$^4$He} 
\def\3he{$^3$He} 
\def\7li{$^7$Li} 
 
\def\yd{$y_{\rm D}$~} 
\def\hii{H\thinspace{$\scriptstyle{\rm II}$}~} 
\def\Nnu{$N_{\nu}$~} 
\def\gsim{\mathrel{\raise.3ex\hbox{$>$\kern-.75em\lower1ex\hbox{$\sim$}}}} 
\newcommand\la{\lower0.6ex\vbox{\hbox{\ensuremath{\buildrel{\textstyle<}\over{\sim}\ }}}} 
\newcommand\ga{\lower0.6ex\vbox{\hbox{\ensuremath{\buildrel{\textstyle>}\over{\sim}\ }}}}

\begin{document}

\title{Effective number of neutrinos and baryon asymmetry from BBN and WMAP} 
 
\author{$^{1,6}$V. Barger, $^2$James P. Kneller, $^1$Hye-Sung Lee, 
$^{3,6}$Danny Marfatia and $^{4,5,6}$Gary Steigman} 
\affiliation{$^1$Department of Physics, University of Wisconsin, Madison, WI 53706} 
\affiliation{$^2$Department of Physics, North Carolina State University, 
Raleigh, NC 27695} 
\affiliation{$^3$Department of Physics, Boston University, Boston, MA 02215} 
\affiliation{$^4$Department of Physics, The Ohio State University, Columbus, 
             OH 43210} 
\affiliation{$^5$Department of Astronomy, The Ohio State University, Columbus, 
             OH 43210} 
\affiliation{$^6$Kavli Institute for Theoretical Physics, University of 
California, Santa Barbara, CA 93106}

\begin{abstract} 
 
We place constraints on the number of relativistic degrees of freedom 
and on the baryon asymmetry at the epoch of Big Bang Nucleosynthesis 
(BBN) and at recombination, using cosmic background radiation (CBR) 
data from the Wilkinson Microwave Anisotropy Probe (WMAP), 
complemented by the 
Hubble Space Telescope (HST) Key Project measurement of the Hubble 
constant, along with the latest compilation of deuterium abundances 
and \hii region measurements of the primordial helium abundance. 
The agreement between the derived values of these key cosmological 
and particle physics parameters at these widely separated (in time 
or redshift) epochs is remarkable.  From the combination of 
CBR and BBN data, we 
find the $2\sigma$ ranges for the effective number of neutrinos \Nnu 
and for the baryon asymmetry (baryon to photon number ratio $\eta$) 
to be 1.7--3.0 and 5.53--6.76 $\times 10^{-10}$, respectively. 
 
\end{abstract} 
 
\pacs{} 
%\vskip 2pc] 
\keywords{Suggested keywords} 
 
\maketitle 
 
\section{Introduction} 
 
 The concordance model of cosmology, with dark energy, dark matter, 
baryons, and three flavors of light neutrinos, provides a consistent 
description of BBN ($\sim 20$ minutes), the CBR ($\sim 380$ Kyr), 
and the galaxy formation epochs of the universe ($\gsim 1$ Gyr). 
The standard model has received recent confirmation from the WMAP 
precision measurements of the CBR temperature and polarization 
anisotropy spectra 
\cite{map}.  However, despite the impressive successes of the standard 
model in describing a wide range of cosmological data, the possibility 
remains that there could be non-standard model contributions to the 
total energy density in the radiation era from additional relativistic 
particles. 
 
 In this paper new constraints are placed on any physics beyond the 
standard model that contributes to the energy density like radiation 
({\it i.e.},~decreases with the expansion of the universe as the 
fourth power of the scale factor, independent of the {\it sign} 
of that contribution).  While such new physics may or may not be 
due to extra relativistic degrees of freedom, it is assumed that 
the non-standard contribution to the energy density may be 
parameterized as such.  Simultaneously, constraints are placed on 
the baryon density at widely differing epochs in the evolution of 
the universe.  The keys to these constraints are the recently 
released measurements of the CBR 
anisotropy spectra by the WMAP collaboration, the most recent 
compilation of high redshift, low metallicity deuterium abundances~\cite{k} 
and $^4$He abundances 
relevant to BBN.

\section{Modified relativistic energy density} 
 
The cosmology of interest here begins when the universe is already 
a few tenths of a second old and the temperature is a few MeV.  At 
such early epochs the total energy density receives its dominant 
contribution from all the relativistic particles present (the 
evolution of the universe is said to be ``radiation-dominated'' 
(RD)).  In the standard cosmology, prior to \epm annihilation, 
these relativistic particles are: photons, \epm pairs and three 
flavors of left-handed (\ie one helicity state) neutrinos (and 
their right-handed antineutrinos). Then, the energy density is 
\be 
\rho_{\rm TOT} = \rho_{\rm R} = \rho_{\gamma} + \rho_{e} + 
3\rho_{\nu} = {43 \over 8}\rho_{\gamma}\,, 
\label{rho0} 
\ee 
where $\rho_{\gamma}$ is the energy density in photons (which 
by today have redshifted to become the CBR photons at a temperature 
of about 2.7 K). 
 
In ``standard'' BBN (SBBN) it is assumed that the neutrinos are fully 
decoupled prior to \epm annihilation and do not share in the energy 
transferred from the annihilating \epm pairs to the CBR photons. 
In this approximation, the photons in the post-\epm annihilation universe 
are hotter than the neutrinos by a factor $T_{\gamma}/T_{\nu} 
= (11/4)^{1/3}$, and the relativistic energy density is 
\be 
\rho_{\rm R} = \rho_{\gamma} + 3\rho_{\nu} = 1.6813\rho_{\gamma}. 
\ee 
 
During the RD epoch the age and the energy density are related by 
${4 \over 3}\rho_{\rm R} t^{2} = 1$ (we have chosen units in which 
$8\pi G = 1$), so that once the particle content ($\rho_{\rm R}$) is 
specified, the age of the universe is known as a function of the CBR 
temperature.  In the standard model, 
\be 
{\rm Pre-\epm annihilation}:~~t~T_{\gamma}^{2} = 0.738~{\rm MeV^{2}~s}, 
\label{ttpre} 
\ee 
\be 
{\rm Post-\epm annihilation}:~~t~T_{\gamma}^{2} = 1.32~{\rm MeV^{2}~s}. 
\label{ttpost} 
\ee 
 
The most straightforward variation of the standard cosmology is 
``extra'' energy contributed 
by new, light (relativistic at BBN) particles ``$X$''.  These might, 
but need not be sterile neutrinos. 
When the $X$ are decoupled, in the sense that they don't share in 
the energy released by \epm annihilation, it is convenient to account 
for the extra contribution to the standard-model energy density by 
normalizing it to that of an ``equivalent" neutrino~\cite{ssg}, 
\be 
\rho_{X} \equiv \Delta N_{\nu}\rho_{\nu} = 
{7 \over 8}\Delta N_{\nu}\rho_{\gamma}. 
\label{deln} 
\ee 
For SBBN \Deln = 0, where $\Delta N_{\nu} \equiv 3 + N_\nu$.   For each 
additional ``neutrino-like" particle (\ie any two-component fermion), 
if $T_{X} = T_{\nu}$, then \Deln = 1; if $X$ is a scalar, \Deln = 4/7. 
However, it may well be that the $X$ have decoupled even earlier in 
the evolution of the universe and have failed to profit from the heating 
when various other particle-antiparticle pairs annihilated (or unstable 
particles decayed).  In this case, the contribution to \Deln from each 
such particle will be $< 1$ ($< 4/7$). 
We emphasize that, in principle, we are 
considering {\it any} term in the energy density which scales like 
$a^{-4}$, where $a$ is the scale factor.  In this sense, the 
modification to the usual Friedman equation due to higher dimensional 
effects, as in the Randall-Sundrum model \cite{rs} (see also, 
\cite{rs1,rs2,rs3,rs4,rs5,rs6,rs7,bratt}), can be included as well. 
An important interest in this latter case is that it permits the possibility 
of a {\it negative} contribution to the radiation density ($\Delta N_{\nu} 
< 0$; $N_{\nu} < 3$). 
 
In the presence of such a modification to the relativistic energy 
density, the pre-\epm annihilation energy density in Eq.~(\ref{rho0}) 
is changed to, 
\be 
(\rho_{\rm R})_{pre} = {43 \over 8}\left(1 + 
{7\Delta N_{\nu} \over 43}\right)\rho_{\gamma}. 
\label{rhoxpre} 
\ee 
Any extra energy density ($\Delta N_{\nu} > 0$) speeds up the expansion 
of the universe so that the right-hand side of the time-temperature 
relation in Eq.~(\ref{ttpre}) is smaller by the square root of the 
factor in parentheses in Eq.~(\ref{rhoxpre}), 
\begin{eqnarray} 
S_{pre} \equiv (t/t')_{pre} & = & \left(1 + {7\Delta N_{\nu} \over 
43}\right)^{1/2} \nonumber \\ & = & (1 + 0.1628\Delta N_{\nu})^{1/2}\,, 
\label{sxpre} 
\end{eqnarray} 
where $t'$ is the age of the universe with the extra energy density. 
In the post-\epm annihilation universe the extra energy density is 
diluted by the heating of the photons, so that, 
\be 
(\rho_{\rm R})_{post} = 1.6813\,(1 + 0.1351\Delta N_{\nu})\rho_{\gamma}, 
\label{rhoxpos} 
\ee 
and 
\be 
S_{post} \equiv (t/t')_{post} = (1 + 0.1351\Delta N_{\nu})^{1/2}. 
\label{sxpos} 
\ee 
This latter expression (Eq.~\ref{sxpos}) is also relevant for the 
modification to the spectrum of temperature fluctuations in the CBR 
(when compared with the standard $N_{\nu} = 3$ case). 
 
\section{~Constraints on $N_\nu$ from the CBR} 
 
\begin{figure}[htbp] 
\begin{center} 
\epsfxsize=3.4in 
\epsfbox{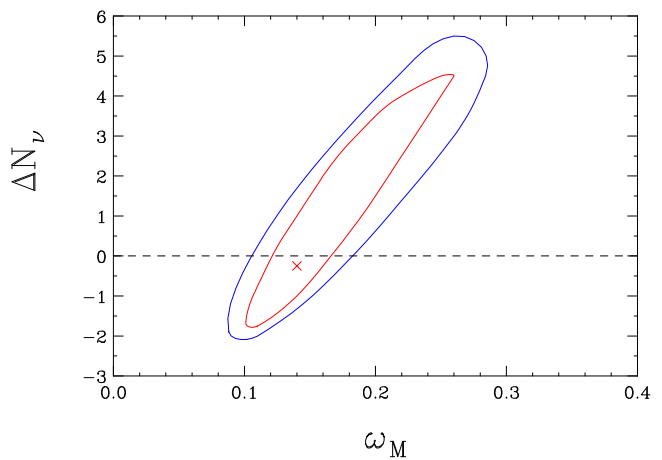} 
      \caption{The CBR degeneracy between $\omega_{\rm M}$ and \Deln 
               is evident from the $1\sigma$ and $2\sigma$ contours 
               from the WMAP data. 
      \label{fig:wmap2}} 
 
\end{center} 
\end{figure}

The competition between gravitational potential and pressure gradients 
is responsible for the peaks and troughs in the CBR power spectrum. The 
redshift of matter-radiation equality, 
\be 
z_{eq}=2.4 \times 10^4\, {\omega_{\rm M} \over S_{post}^2}\,, 
\ee 
affects the time (redshift) duration over which this competition occurs. 
Here, $\omega_{\rm M} \equiv \Omega_{\rm M} h^2$ 
is the total matter density (comprised, for nearly massless neutrinos, 
of baryons and cold dark matter) and $h$ ($H_0 \equiv 100h$ km/s/Mpc) is 
the normalized Hubble constant. 
The direct correlation 
between $\omega_{\rm M}$ and \Deln is evident in Fig.~\ref{fig:wmap2} which 
results from our analysis described below. 
The primary effects of 
relativistic degrees of freedom (other than photons) on the CBR power 
spectrum result essentially from changing the redshift of matter-radiation 
equality. If the radiation content is increased, matter-radiation equality 
is delayed, and occurs closer (in time and/or redshift) to the epoch of 
recombination. 
 
The redshift of matter-radiation equality is important for 
two reasons~\cite{hu}: 
\begin{itemize} 
\item {Radiation causes potential decay which blueshifts the photons 
because they do not have to climb out of such deep wells.  Moreover, 
the concurrent decay in the spatial curvature doubles the blueshift 
effect by contracting the wavelength of the photons relative to the 
pure cosmological expansion.} 
\item{In the matter dominated (MD) era before recombination, the 
density contrast ($\delta \rho/\rho$) of the {\it pressureless} cold 
dark matter (CDM) grows unimpeded (as $t^{2/3}$) while the density 
contrast of the baryons is either oscillating or decaying. The longer 
this pre-recombination MD era lasts, the more suppressed are the 
amplitudes of the peaks.} 
\end{itemize} 
Conversely, if matter-radiation equality is delayed, the gravitational 
potential is dominated by the photon-baryon fluid closer to recombination 
resulting in a more pronounced peak structure.  
 
\begin{figure}[htbp] 
\begin{center} 
\epsfxsize=3.4in 
\epsfbox{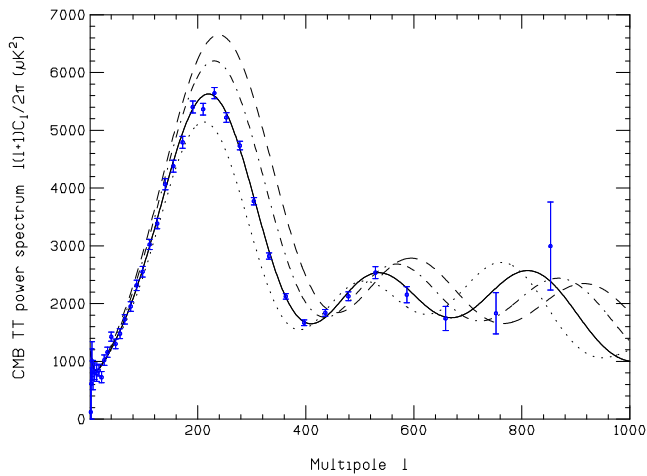 } 
      \caption{The power spectrum for the best-fit (\Nnu$=2.75$) to 
the WMAP data is the solid line. With all other parameters and the 
overall normalization of the primordial spectrum fixed, 
the spectra for \Nnu$=1$, \Nnu$=5$ and 
\Nnu$=7$ are the dotted, dot-dashed and dashed lines, respectively. The data 
points represent the binned TT power spectrum from WMAP. 
      \label{fig:spectrum}} 
 
\end{center} 
\end{figure}

An increase in the relativistic content causes the universe to be younger 
at recombination with a correspondingly smaller sound horizon $s_*$. Since 
the location of the $n^{th}$ peak scales roughly as $n\pi D_*/s_*$ (where 
$D_*$ is the comoving angular diameter distance to recombination), the 
peaks shift to smaller angular scales (larger $l$) and with greater 
separation. These features are clearly visible in 
Fig.~\ref{fig:spectrum}. 

The heights and locations of the peaks also depend on the 
history of the universe after recombination. 
At the end of matter domination and the onset of dark energy domination, 
further and much slower (compared to that in the radiation epoch) 
potential decay occurs.  The more gradual potential decay causes the 
induced anisotropy to be suppressed by a factor of $l$. The amplification
of the power in the lowest $l$'s from this late decay serves 
as a probe of dark energy 
(or another probe of the matter content in a flat universe). 
In principle, 
the degeneracy between $\Delta N_{\nu}$ and $\omega_{\rm M}$  
is broken by this effect and by the accompanying change 
in the redshift at which 
the matter dominated epoch ends. 
However, note that the lowest multipoles also 
have the largest cosmic variance. 

The TT and TE power spectra are computed using the Code for Anisotropies 
in the Microwave Background or CAMB~\cite{camb} which is a parallelized 
version of CMBFAST~\cite{cmbfast}.  The Universe is  assumed to be flat, 
in accord with the predictions of inflation~\cite{inflation}, and the 
dark energy is assumed to behave as a cosmological constant $\Lambda$. 
The restriction of a flat geometry allows us to relate 
the dark energy and matter densities at the present time: 
$\Omega_{\Lambda} = 1 - \Omega_{M}$. 
The angular power spectrum is calculated on a grid defined by $h$, the 
baryon density $\omega_{\rm B} \equiv \Omega_{\rm B} h^2$ (or $\eta_{10} 
\equiv 10^{10}n_{\rm B}/n_{\gamma} = 274\omega_{\rm B}$), $\omega_{\rm M}$, 
the number of equivalent neutrinos $N_{\nu}$, the reionization optical depth 
$\tau$, and the spectral index $n_s$ of the primordial power spectrum. 
Two priors are imposed to largely break the degeneracy between 
$\omega_{\rm M}$ and 
\Deln.  For $h$ a top-hat distribution is chosen corresponding to the 
HST measurement, $h=0.72\pm 0.08$~\cite{HST}, and we require that the 
universe be older than the globular clusters (which, at $2\sigma$, are 
older than 11 Gyr~\cite{chaboyer}).  For comparison, we also consider 
the case when the age of the universe $t_0$ exceeds 12 Gyr.

Our top-hat grid, consisting of over 10 million points, is: 
\begin{itemize} 
\item {$0.64 \leq h \leq 0.8$ in steps of size 0.02.} 
\item {$0.018 \leq \omega_{\rm B} \leq 0.028$ in steps of size 0.001.} 
\item {$0.11 \leq \omega_{\rm M} \leq 0.27$ in steps of size 0.01 and 
$\omega_{\rm M}=0.07$, 0.3.} 
\item{$1 \leq$ \Nnu $\leq 3.5$ in steps of size 0.25, $4 \leq$ \Nnu 
$\leq 9$ in steps of size 0.5 and \Nnu = 0, 0.5.} 
\item {$0 \leq \tau \leq 0.3$ in steps of size 0.025.} 
\item{$0.90 \leq n_s \leq 1.02$ in steps of size 0.01 and $n_s=0.80$, 0.84, 
0.88, 1.04, 1.08, 1.12, 1.16, 1.20.} 
\item{The normalization of the spectrum is a continuous parameter.} 
\end{itemize} 
 
The first year WMAP data are in the form of 899 measurements of the TT power 
spectrum from $l=2$ to $l=900$~\cite{maptt} and 449 data points 
of the TE power spectrum~\cite{mapte}.  Although the effect of 
relativistic degrees of freedom on the TE spectrum is insignificant, 
it is included in our 
analysis for completeness.  The likelihood of each model of our 
grid is computed using Version 1.1 of the code provided by the WMAP 
collaboration~\cite{mapcode}. The code computes the covariance matrix 
under the assumption that the off-diagonal terms are subdominant. 
This approximation breaks down for unrealistically small amplitudes. 
When the height of the first peak is below 5000 $\mu K^2$ (which is 
many standard deviations away from the data), only the diagonal terms 
of the covariance matrix are used to compute the likelihood. 
 
\begin{figure}[htbp] 
\begin{center} 
\epsfxsize=3.4in 
\epsfbox{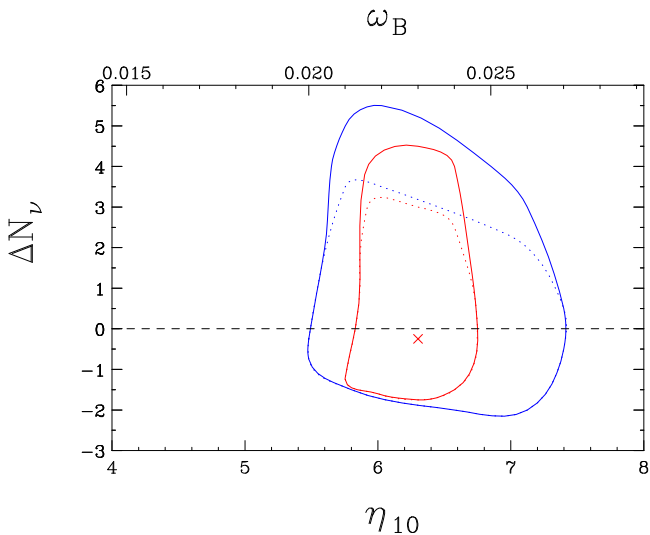} 
      \caption{The $1\sigma$ and $2\sigma$ contours in the 
               $\eta_{10}$--\Deln plane from WMAP data. The 
solid (dotted) lines correspond to $t_0>11$ (12) Gyr. The cross marks 
the best-fit at $\omega_{\rm B}=0.023$ and \Deln$=-0.25$. 
      \label{fig:wmap}} 
 
\end{center} 
\end{figure} 
 
The best-fit parameters are $h=0.68$, $\omega_{\rm B}=0.023$ 
($\eta_{10} = 6.3$), 
$\omega_{\rm M}=0.14$, \Nnu$=2.75$, $\tau=0.13$, and $n_s=0.97$ 
with a $\chi^2=1429.13$ for 1341 degrees of freedom. The allowed 
parameter space in the $\eta_{10}$--\Deln plane is shown in 
Fig.~\ref{fig:wmap}. The solid (dotted) lines correspond to the 
$1\sigma$ and $2\sigma$ regions{\footnote{For 2-dimensional 
constraints, the $1\sigma$, $2\sigma$ and $3\sigma$ regions are 
defined by $\Delta \chi^2=2.3$, 6.17 and 11.83, respectively.}} 
for $t_0>11$ (12) Gyr.  The cross identifies the best-fit 
point. Note that while this best fit point lies at \Nnu $< 3$, 
the \Nnu distribution is very broad.  After marginalizing over 
$\eta$, the $2\sigma$ range in \Nnu extends from 
0.9 (\Deln = $-2.1$) to 8.3 (\Deln = 5.3).  Although similar CBR 
analyses (see Ref.~\cite{others}) have included different, additional 
data to that from WMAP alone, making direct comparisons difficult, 
our results are in good agreement with them.  The prior on $t_0$ 
has a significant effect on the allowed values of \Deln\cite{kssw} for a 
simple reason. Since flatness is assumed, $t_0$ depends only on 
the matter content and the Hubble parameter via 
\begin{equation} 
H_{0}\,t_{0} \simeq {2\over 3} {1\over \sqrt{1-\Omega_{\rm M}}} 
\ln\bigg({1+\sqrt{1-\Omega_{\rm M}}\over \sqrt{\Omega_{\rm M}}}\bigg). \label{eq:age} 
\end{equation} 
The combination of the HST prior on 
$h$ and the $t_0$ prior restricts $\omega_{\rm M}$ and help to break the 
degeneracy between \Deln and $\omega_{\rm M}$. 
 
The best fit WMAP-determined baryon density is $\eta_{10} = 6.30$ 
($\omega_{\rm B} = 0.0230$), in excellent agreement with Spergel 
\etal \cite{sperg} and other similar analyses~\cite{others}.  The CBR 2$\sigma$ 
range extends from $\eta_{10} = 5.58$ ($\omega_{\rm B} = 0.0204$) 
to $\eta_{10} = 7.26$ ($\omega_{\rm B} = 0.0265$). 
 
These CBR constraints on \Nnu and $\omega_{\rm B}$ apply to epochs 
in the evolution of the universe $\ga 380$ Kyr.  An important test 
of the standard models of cosmology and particle physics is to compare 
them with corresponding constraints from the much earlier epoch probed 
by BBN. 
 
\section{The roles of \Nnu and $\eta_{10}$ in BBN} 
 
At $T \sim$ few MeV, the neutrinos are beginning to decouple from the 
$\gamma$ -- \epm plasma and the neutron to proton ratio, crucial for 
the production of primordial \4he, is decreasing. As the temperature 
drops below $\sim 2$~MeV, the two-body collisions between neutrinos 
and \epm pairs, responsible for keeping the neutrinos in thermal 
equilibrium with the electron-positron -- photon plasma become slow 
compared to the universal expansion rate and the neutrinos decouple, 
although they do continue to interact with the neutrons and protons 
via the charged-current weak interactions. Prior to \epm annihilation, 
when the temperature drops below $\sim 0.8$~MeV and the universe is 
$\approx 1$~second old, these interactions, interconverting neutrons 
and protons, become too slow (compared to the universal expansion rate) 
to maintain $n-p$ equilibrium and the neutron-to-proton ratio begins 
to deviate from ({\it exceeds}) its equilibrium value ($(n/p)_{eq} 
= \exp(-\Delta m/T)$), where $\Delta m$ is the neutron-proton mass 
difference.  Beyond this point, often described as neutron-proton 
``freeze-out'', the $n/p$ ratio continues to decrease, albeit more 
slowly than would have been the case in equilibrium.  Since there 
are several billion CBR photons for every nucleon (baryon), the 
abundances of any complex nuclei are entirely negligible at these 
early times. 
 
We note here that if there is an {\it asymmetry} between the numbers 
of $\nu_{e}$ and $\bar\nu_{e}$ (``neutrino degeneracy''), described 
by a chemical potential $\mu_{e}$, then the equilibrium neutron-to-proton 
ratio is modified to $(n/p)_{eq} = \exp(-\Delta m/T - \mu_{e}/T)$. 
In place of the neutrino chemical potential, it is convenient to 
introduce the dimensionless degeneracy parameter $\xi_{e} \equiv 
\mu_{e}/T$.  A positive chemical potential ($\xi_{e} 
> 0$; more $\nu_{e}$ than $\bar\nu_{e}$) leads to {\it fewer} 
neutrons and less \4he will be synthesized in BBN. 
 
BBN begins in earnest {\it after} \epm annihilation, at $T \approx 
0.08$~MeV ($t \approx 3$~minutes), when the number density of those CBR 
photons with sufficient energy to photodissociate deuterium (those in the 
tail of the black body distribution) is comparable to the baryon density. 
By this time the $n/p$ ratio has further decreased (the two-body reactions 
interconverting neutrons and protons having been somewhat augmented by 
ordinary beta decay; $\tau_{n} = 885.7$~sec.), limiting (mainly) the amount 
of helium-4 which can be synthesized.  As a result, the predictions of the 
primordial abundance of \4he  depend sensitively on the early expansion 
rate and on the amount -- if any -- of a $\nu_{e} - \bar\nu_{e}$ asymmetry. 
 
In contrast to \4he, the BBN-predicted abundances of deuterium, helium-3 
and lithium-7 (the most abundant of the nuclides synthesized during BBN) 
are determined by the competition between the various two-body 
production/destruction rates and the universal expansion rate.  As a 
result, the D, \3he, and \7li abundances are sensitive to the post-\epm 
annihilation expansion rate, while that of \4he depends on {\it both} the 
pre- and post-\epm annihilation expansion rates; the former determines 
the ``freeze-in" and the latter modulates the importance of beta decay 
(see, \eg Kneller \& Steigman \cite{knellsteig}).  Also, the primordial 
abundances of D, \3he, and Li, while not entirely insensitive to neutrino 
degeneracy, are much less effected by a non-zero $\xi_{e}$ (\eg \cite{ks}). 
 
Of course, the BBN abundances do depend on the baryon density which fixes 
the nuclear reactions rates and also, through the {\it ratio} of baryons 
to photons, regulates the time/temperature at which BBN begins.  As a result, 
the abundances of at least two different relic nuclei are needed to break 
the degeneracy between the baryon density and a possible non-standard 
expansion rate resulting from new physics or cosmology, and/or a neutrino 
asymmetry.  In this paper only the former possibility is considered; in 
another publication several of us (along with P. Langacker) have explored 
the consequences of neutrino degeneracy and we studied the modifications 
to the constraints on \Deln when both of these non-standard effects are 
 simultaneously included. 
 
\begin{figure}[htbp] 
\begin{center} 
\epsfxsize=3.4in 
\epsfbox{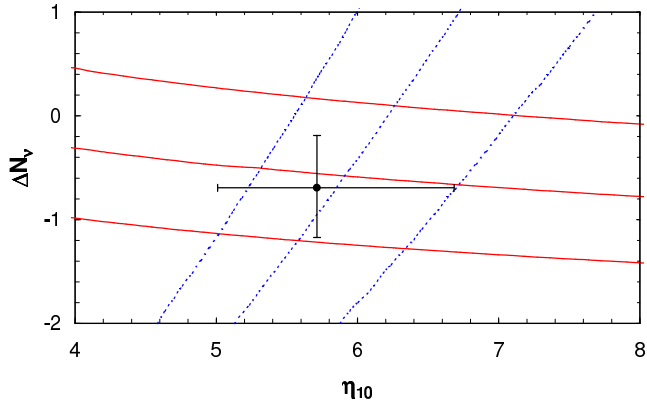} 
      \caption{Isoabundance curves for D and \4he in the 
      $\eta_{10}$ -- \Deln plane.  The solid curves are 
      for  \4he (from top to bottom: Y = 0.25, 0.24, 0.23). 
      The dashed curves are for D (from left to right: 
      $10^{5}$(D/H) = 3.0, 2.5, 2.0.  The data point with 
      error bars corresponds to \yd = 2.6$\pm 0.4$ and Y 
      = 0.238$\pm 0.005$; see the text for discussion of 
      these abundance values. 
      \label{bbnf1}} 
 
\end{center} 
\end{figure} 
 
While the abundances of D, $^3$He, and Li are most sensitive to the 
baryon density ($\eta$), the $^4$He mass fraction (Y) provides the 
best probe of the expansion rate.  This is illustrated in Fig.~\ref{bbnf1} 
where, in the \Deln -- $\eta_{10}$ plane, are shown isoabundance 
contours for D/H and Y (the isoabundance curves for $^3$He/H and for 
Li/H, omitted for clarity, are similar in behavior to that of D/H). 
The trends illustrated in Fig.~\ref{bbnf1} are easy to understand in 
the context of the discussion above.  The higher the baryon density 
($\eta_{10}$), the faster primordial D is destroyed, so the relic 
abundance of D is {\it anticorrelated} with $\eta_{10}$.  But, the 
faster the universe expands (\Deln $> 0$), the less time is available 
for D-destruction, so D/H is positively, albeit weakly, correlated 
with $\Delta N_\nu$.  In contrast to D (and to \3he and Li), since 
the incorporation of all available neutrons into \4he is not limited 
by the very rapid nuclear reaction rates, the \4he mass fraction is 
relatively insensitive to the baryon density, but it is very sensitive 
to both the pre- and post-\epm annihilation expansion rates (which 
control the neutron-to-proton ratio).  The faster the universe expands, 
the more neutrons are available for \4he.  The very slow increase of 
Y with $\eta_{10}$ is a reflection of the fact that for higher baryon 
density, BBN begins earlier, when there are more neutrons.  As a result 
of these complementary correlations, the pair of primordial abundances 
$y_{\rm D} \equiv 10^{5}(D/H)$ and the \4he mass fraction Y, provide 
observational constraints on both the baryon density and the universal 
expansion rate when the universe was some 20 minutes old.  Comparing 
these to constraints when the universe was some 380 Kyr old, 
from the WMAP observations of the CBR 
spectra, provides a test of the consistency 
of the standard models of cosmology and of particle physics and further 
constrains the allowed range of the present baryon density of the universe.

\section{~Primordial Abundances} 
 
It is clear from Fig.~\ref{bbnf1} that while D (and/or \3he and/or 
\7li) largely constrains the baryon density and \4he plays the same 
role for $\Delta N_\nu$, there is an interplay between $\eta_{10}$ 
and \Deln which is quite sensitive to the adopted abundances.  For 
example, a {\it lower} primordial D/H {\it increases} the BBN-inferred 
value of $\eta_{10}$, leading to a {\it higher} predicted primordial 
\4he mass fraction.  If the primordial \4he mass fraction derived 
from the data is ``low", then a low upper bound on \Deln will be 
inferred.  Therefore, it is crucial to make every effort to avoid 
biasing any conclusions by {\it underestimating} the present 
uncertainties in the primordial abundances derived from the 
observational data.  For this reason deuterium is adopted as the 
baryometer of choice.  Primarily, this is because its observed 
abundance should have only decreased since BBN \cite{els}, but 
also because the deuterium observed in the high redshift, low 
metallicity QSO absorption line systems (QSOALS) should be very 
nearly primordial.  In contrast, the post-BBN evolution of \3he 
and of \7li are considerably more complicated, involving competition 
between production, destruction, and survival.  As a result, at least 
so far, the current, locally observed (in the Galaxy) abundances of 
these nuclides have been of less value in constraining the baryon 
density than has deuterium.  Nonetheless, inferring the primordial 
D abundance from the QSOALS has not been without its difficulties, 
with some abundance claims having been withdrawn or revised. 
Presently there are 5 -- 6 QSOALS with reasonably firm deuterium 
detections \cite{bta,btb,o'm,pb,dod,k}.  However, there is 
significant dispersion among the abundances and the data fail 
to reveal the anticipated ``deuterium plateau" at low metallicity 
or at high redshift \cite{gs01}.  Furthermore, subsequent observations 
of the D'Odorico \etal \cite{dod} QSOALS by Levshakov \etal 
\cite{lev01} revealed a more complex velocity structure and 
led to a revised -- and more uncertain -- deuterium abundance.  This 
sensitivity to the often poorly constrained velocity structure 
in the absorbers is also exposed by the analyses of published 
QSOALS data by Levshakov and collaborators \cite{lev1,lev2,lev3}, 
which lead to consistent, but somewhat higher deuterium abundances 
than those inferred from ``standard" data reduction analyses. 
In the absence of a better motivated choice, here we adopt the 
five abundance determinations collected in the recent paper of 
Kirkman \etal \cite{k}.  The weighted mean value of \yd is 2.6~\footnote 
{This differs from the result quoted in Kirkman \etal 
because they have taken the mean of log($y_{\rm D}$) and then 
used it to infer \yd (\yd $ \equiv 10^{<log(y_{\rm D})>}$).}. 
But, the dispersion among these five data points is very large. 
For this data set $\chi^{2} = 15.3$ for four degrees of freedom, 
suggesting that one or more of these abundance determinations may 
be in error, perhaps affected by unidentified and unaccounted for 
systematic errors.  For this reason, we follow the approach advocated 
by \cite{o'm} and \cite{k} and adopt for the uncertainty in \yd 
the dispersion divided by the square root of the number of data 
points.  Thus, the primordial abundance of deuterium to be used 
 here is chosen to be: \yd $ = 2.6 \pm 0.4$.  For SBBN ($N_\nu 
= 3$, $\xi_{e} = 0$), at $\pm 1\sigma$ this corresponds to a baryon 
density $\eta_{10} = 6.1^{+0.7}_{-0.5}$ ($\omega_{\rm B} = 0.022 
\pm 0.002$)~\footnote{We have purposely avoided quoting the baryon 
density to more significant figures than is justified by the accuracy 
of the D-abundance determination.}. 
 
A similar, less than clear situation exists for determinations of 
the primordial abundance of \4he.  At present there are two, largely 
independent, estimates based on analyses of large data sets of 
low-metallicity, extragalactic \hii regions.  The ``IT" \cite{itl,it} 
estimate of Y(IT) $= 0.244 \pm 0.002$, and the ``OS" determination 
\cite{os,oss,fo} of Y(OS) $= 0.234 \pm 0.003$ which differ by nearly 
$3\sigma$.  The recent analysis of high quality observations of a 
relatively metal-rich (hence, chemically evolved and post-primordial) 
\hii region in the Small Magellanic Cloud (SMC) by Peimbert, Peimbert, 
and Ruiz (PPR) \cite{ppr} yields an abundance Y$_{\rm SMC} = 0.2405 
\pm 0.0018$.  When PPR extrapolated this abundance to zero metallicity, 
they found Y(PPR) $= 0.2345 \pm 0.0026$, lending support to the OS 
value. These comparisons of different observations and analyses 
suggest that unaccounted systematic errors may dominate the statistical 
uncertainties.  Indeed, Gruenwald, Steigman, and Viegas \cite{gsv} 
argue that if unseen neutral hydrogen in the ionized helium region 
of the observed \hii regions is accounted for, the IT estimate of 
the primordial abundance should be reduced to Y(GSV) $= 0.238 \pm 
0.003$ (see also \cite{vgs,sj}).  Here, we adopt this latter estimate 
for the central value but, as we did with deuterium, the uncertainty 
is increased in an attempt to account for likely systematic errors: 
Y $= 0.238 \pm 0.005$, leading to a 2$\sigma$ range, $0.228 
\le $~Y $ \le 0.248$; this range is in accord with the estimate 
adopted by Olive, Steigman, and Walker (OSW) \cite{osw} in their 
review of SBBN.  Although we will comment on the modification to 
any conclusions if Y(IT) is substituted for Y(OSW), 
Figs.~\ref{bbnf1} -- \ref{fig:wmap+bbn} are shown for 
$y_{\rm D} = 2.6 \pm 0.4$ and Y(OSW) = $0.238 \pm 0.005$.

\section{Standard BBN} 
 
\begin{figure}[htbp] 
\begin{center} 
\epsfxsize=3.4in 
\epsfbox{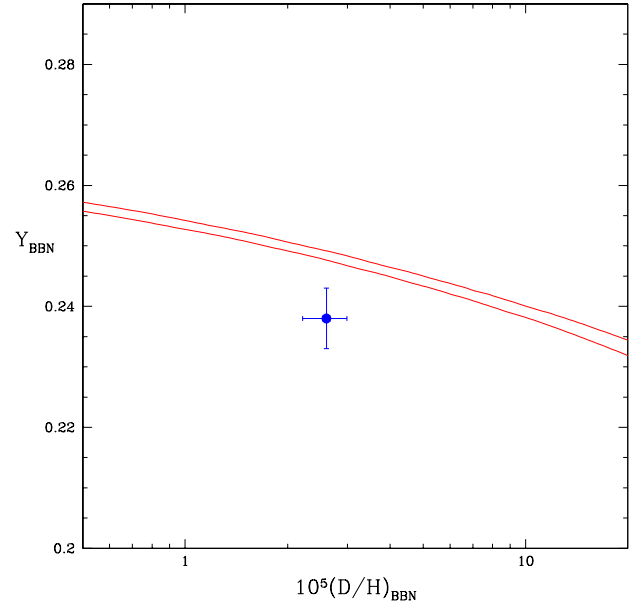} 
      \caption{The band is the SBBN predicted relation between the 
      primordial abundances of D and \4he, including the errors 
      ($\pm 1\sigma$) in those predictions from the uncertainties 
      in the nuclear and weak interaction rates.  The point with 
      error bars is for the relic abundances of D and \4he adopted 
      here (see the text). 
      \label{bbnf3}} 
 
\end{center} 
\end{figure} 
 
Before proceeding to our main goal of constraining new physics using 
BBN, it is worthwhile to set the scene by considering the standard 
model case ($N_\nu$ = 3, $\xi_{e} = 0$) first.  The result of this 
comparison is well known: there is a ``tension'' between the primordial 
abundances of D and \4he inferred from the observational data and 
those predicted by SBBN.  For example, if \yd is used to fix the 
baryon density (\yd $ = 2.6 \pm 0.4$) then, at $\pm 1\sigma$, 
$\eta_{10}^{\rm SBBN} = 6.1^{+0.7}_{-0.5}$ ($\omega_{\rm B} = 
0.022 \pm 0.002$), the corresponding predicted \4he abundance is 
Y $ = 0.248 \pm 0.001$ ($1\sigma$), which is some 2$\sigma$ higher 
than either the OSW or the IT estimates.  This is illustrated in 
Fig.~\ref{bbnf3} which shows the SBBN-predicted relation between the 
relic abundances of D and \4he along with the Y(OSW) abundance estimate 
adopted here.  The D-inferred baryon density is in 
excellent agreement with the baryon density determined independently 
(non-BBN; $N_\nu = 3$) by Spergel \etal \cite{sperg} from a combination 
of CBR and Large Scale Structure data (2dF + Lyman $\alpha$): 
$\eta_{10}^{\rm non-BBN} = 6.14 \pm 0.25$ ($\omega_{\rm B} = 0.0224 
\pm 0.0009$)~\footnote{It should be noted that from the CBR {\it alone} 
Spergel \etal find $\eta_{10}^{\rm CBR} = 6.6 \pm 0.3$ ($\omega_{\rm B} 
= 0.024 \pm 0.001$).  It is this value which is most directly comparable 
to our CBR and joint BBN/CBR results.}.  Thus, it appears that \4he is 
the problem: the primordial abundance of \4he is smaller than predicted 
for SBBN given {\it either} the observed deuterium abundance {\it or} 
the non-BBN inferred baryon density.  At the same time we strongly 
emphasize that this ``discrepancy'' is only at the $\sim 2\sigma$ level 
and we should celebrate the excellent agreement between the baryon 
density determined when the universe was only 20 minutes old and when 
the universe was some 380 Kyr old (CBR). 
 
\section{Non-Standard BBN: \Deln $ \neq 0$} 
 
\begin{figure}[htbp] 
\begin{center} 
\epsfxsize=3.4in 
\epsfbox{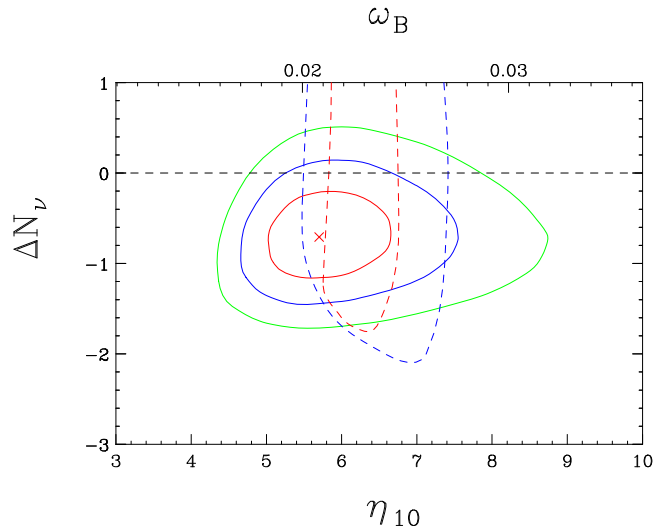} 
      \caption{The $1\sigma$, $2\sigma$ and $3\sigma$ contours in 
               the $\eta_{10}$--\Deln plane for the adopted D and 
               \4he (OSW) abundances (solid lines).  The cross 
               marks the best fit BBN point.  The $1\sigma$ and 
               $2\sigma$ contours from WMAP (dashed lines) are 
               shown for comparison. 
      \label{bbnf4}} 
 
\end{center} 
\end{figure}

As noted above, for SBBN ($N_\nu = 3$) the observationally 
inferred primordial abundance of \4he is too small (by $\sim 
2\sigma$) for the baryon density inferred {\it either} from 
the D abundance {\it or} from the non-BBN analysis of Spergel 
\etal \cite{sperg}.  This suggests that the early universe 
expansion rate may have been too fast, leaving too many neutrons 
available for the synthesis of \4he.  If this tension between 
\4he and D (or, between \4he and $\omega_{\rm B}$) should persist, 
it could be a signal of non-standard physics corresponding to 
$S < 1$ (\Deln $< 0$).  Indeed, in Fig.~\ref{bbnf1} it can be 
seen that for the adopted primordial abundances of D and \4he, 
there is a ``perfect'' fit ($\chi^{2} = 0$) for \Deln $ \approx 
-0.7$ ($N_\nu \approx 2.3$) and  $\eta_{10} \approx 5.7$. 
Although $N_\nu = 3$ is only disfavored by $\sim 2\sigma$, 
any {\it increase} in the early universe expansion rate ($S > 
1$, $N_\nu > 3$) is strongly disfavored.  This is illustrated 
in Fig.~\ref{bbnf4} which shows the 1$\sigma$, 2$\sigma$ 
and 3$\sigma$ contours in the $\eta_{10}$--\Deln plane for 
the adopted D and \4he (OSW) abundances.  The shape of these 
contours reflects our much discussed complementarity between 
D and \4he: D provides the best constraint on the baryon 
density while \4he is most sensitive to the early universe 
expansion rate, the latter providing an excellent probe of possible 
new physics. 
 
With reference to that Fig.~\ref{bbnf4} we note that even one 
extra, fully thermalized neutrino (\Deln = 1) is strongly 
disfavored.  This seemingly eliminates the sterile neutrino 
suggested by the LSND experiment~\cite{lsnd}.  For the LSND 
parameters, in the absence of significant neutrino asymmetry, 
this ``sterile'' neutrino would have been mixed with the 
active neutrinos and thermalized prior to neutrino decoupling 
(prior to BBN)~\cite{paul}.  If such a neutrino were to exist, the ``new'' 
standard model would correspond to \Nnu = 4.  This would be a 
disaster since for \Nnu = 4 and the OSW \4he abundance the 
{\it minimum} $\chi^{2}$ (which now occurs at $\eta_{10} \approx 
6.1$) is greater than 20.  The situation is even worse 
for the IT \4he abundance since the smaller uncertainty in Y forces 
a much smaller baryon density   
($\eta_{10} \approx 4.7$), with $\chi^{2}_{min} > 60$! 
 
\subsection{Requiring $N_{\nu} \geq ~3$} 
 
\begin{figure}[htbp] 
\begin{center} 
\epsfxsize=3.4in 
\epsfbox{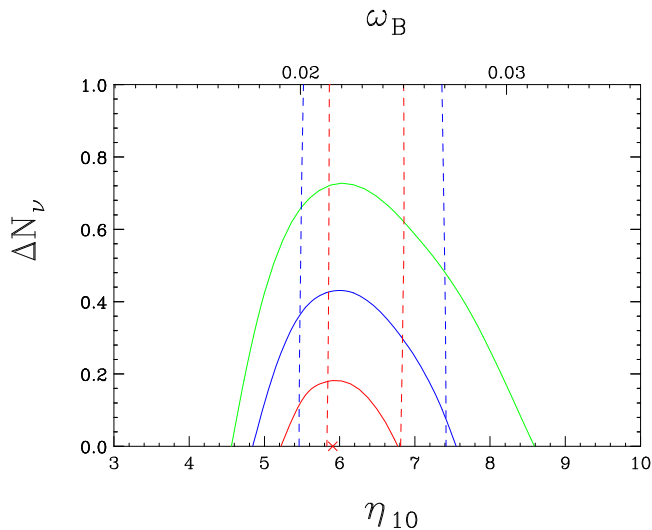} 
      \caption{The $1\sigma$, $2\sigma$ and $3\sigma$ contours 
               (solid lines) in the $\eta_{10}$--\Deln plane 
               for \Nnu $ \geq 3$ and the adopted D and \4he 
               (OSW) abundances.  The corresponding $1\sigma$ and $2\sigma$ 
               contours from WMAP (dashed lines) are shown for 
               comparison. 
      \label{bbnf5}} 
 
\end{center} 
\end{figure}

Since, as is well known from LEP~\cite{pdg}, there are three flavors 
of active, left-handed neutrinos (and their right-handed antiparticles), 
any {\it extra} contributions to the relativistic energy density 
at BBN should result in $N_\nu > 3$.  Of course, ``new physics'' 
in the form of non-minimally coupled fields (\eg \cite{css,knellsteig} 
and references therein) or from higher-dimensional phenomena (\eg 
\cite{rs,rs1,rs2,rs3,rs4,rs5,rs6,rs7,bratt}) may result in an {\it 
effective} $N_\nu < 3$.  If, however, the class of non-standard 
physics of interest is restricted to \Deln $ \geq 0$, then the BBN 
constraints presented above (and those from the CBR) will change. 
With a prior of $N_\nu \geq 3$, the best fit BBN-determined values 
of the baryon-to-photon ratio and \Deln (for Y(OSW)) shift from 
$\eta_{10} \approx 5.7$ and $\Delta$$N_\nu \approx -0.7$, to 
$\eta_{10} \approx 5.9$ and $\Delta$$N_\nu = 0$.  The value 
of $\chi^{2}_{min}$ changes to 4.2.  The corresponding confidence contours 
in the $\eta_{10}$--\Deln plane are shown in Fig.~\ref{bbnf5}. 
 
\section{Joint Constraints And Summary} 
 
\begin{figure}[htbp] 
\begin{center} 
\epsfxsize=3.4in 
\epsfbox{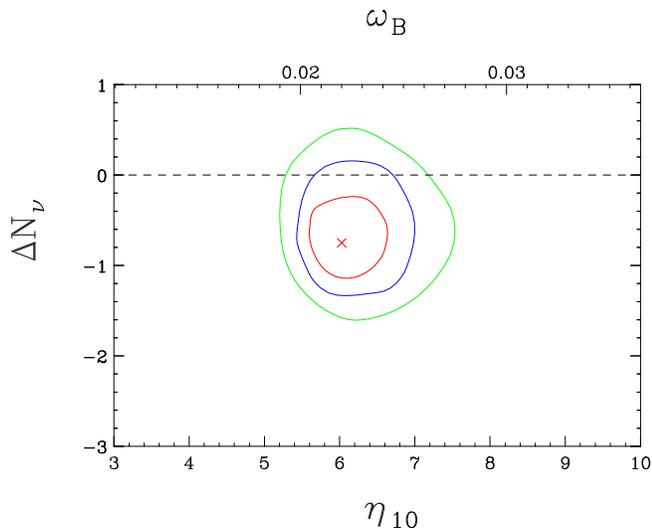} 
      \caption{The $1\sigma$, $2\sigma$ and $3\sigma$ contours in the 
               $\eta_{10}$--\Deln plane from a combination of WMAP data 
               and the adopted D and \4he (OSW) abundances. 
      \label{fig:wmap+bbn}} 
 
\end{center} 
\end{figure} 
 
As may be seen from Figs.~\ref{bbnf4} and \ref{bbnf5}, the agreement 
between the values obtained for \Deln and $\eta_{10}$ from WMAP and 
from BBN {\it separately} is excellent.  Guided by this, the BBN and CBR 
results are combined to obtain the joint fit in the $\eta$ -- \Deln 
plane shown in Fig.~\ref{fig:wmap+bbn}.  To a good first approximation, 
BBN (and primordial \4he) determines \Deln while WMAP fixes $\eta_{10}$ 
(with some help from BBN and primordial D).  The corresponding figure 
for \Nnu $\geq 3$ for the joint fit has not been shown because again the 
\Deln range is almost identical to that from BBN alone (Fig.~\ref{bbnf5}). 
 
\begin{table}[h] 
\begin{center} 
\begin{tabular}{|l|c|c|} 
\hline 
  &\Nnu (2$\sigma$ range)  & $\eta_{10}$ (2$\sigma$ range) \\ \hline 
  WMAP& 0.9 -- 8.3 & 5.58 -- 7.26\\ \hline 
  \yd + Y(OSW) & 1.7 -- 3.0 & 4.84 -- 7.11\\ \hline 
  \yd + Y(IT) & 2.4 -- 3.0 & 5.06 -- 7.33\\ \hline 
  WMAP + \yd + Y(OSW)&1.7 -- 3.0 & 5.53 -- 6.76\\ \hline 
  WMAP + \yd + Y(IT) &2.4 -- 3.0 & 5.58 -- 6.71\\ \hline 
\end{tabular} 
\label{tab1} 
\end{center} 
\caption{The 2$\sigma$ ranges (for 1 degree of freedom) of \Nnu and 
$\eta_{10}$ from analyses of WMAP data, deuterium and helium abundances 
and their combinations. The WMAP analysis involves the assumption of 
a flat universe, along with the strong HST prior on $h$ and the age constraint 
$t_0>11$ Gyr. For BBN the adopted primordial abundances are: \yd $\equiv 
10^{5}$(D/H)$=2.6 \pm 0.4$, Y(OSW) = $0.238 \pm 0.005$, and Y(IT)$= 
0.244 \pm 0.002$.} 
\end{table} 

\begin{table}[h] 
\begin{center} 
\begin{tabular}{|l|c|c|} 
\hline 
  &\Nnu (2$\sigma$ bound)  & $\eta_{10}$ (2$\sigma$ range) \\ \hline 
  WMAP& 8.3 & 5.64 -- 7.30\\ \hline 
  \yd + Y(OSW) & 3.3 & 5.04 -- 7.18\\ \hline 
  \yd + Y(IT) & 3.1 & 4.89 -- 6.56\\ \hline 
  WMAP + \yd + Y(OSW)& 3.3 & 5.66 -- 6.80\\ \hline 
  WMAP + \yd + Y(IT) &3.1  & 5.54 -- 6.60\\ \hline 
\end{tabular} 
\label{tab2} 
\end{center} 
\caption{The same as Table~I, except that the constraint \Nnu $\geq 3$ 
is imposed. } 
\end{table} 
 
\begin{figure}[htbp] 
\begin{center} 
\epsfxsize=3.4in 
\epsfbox{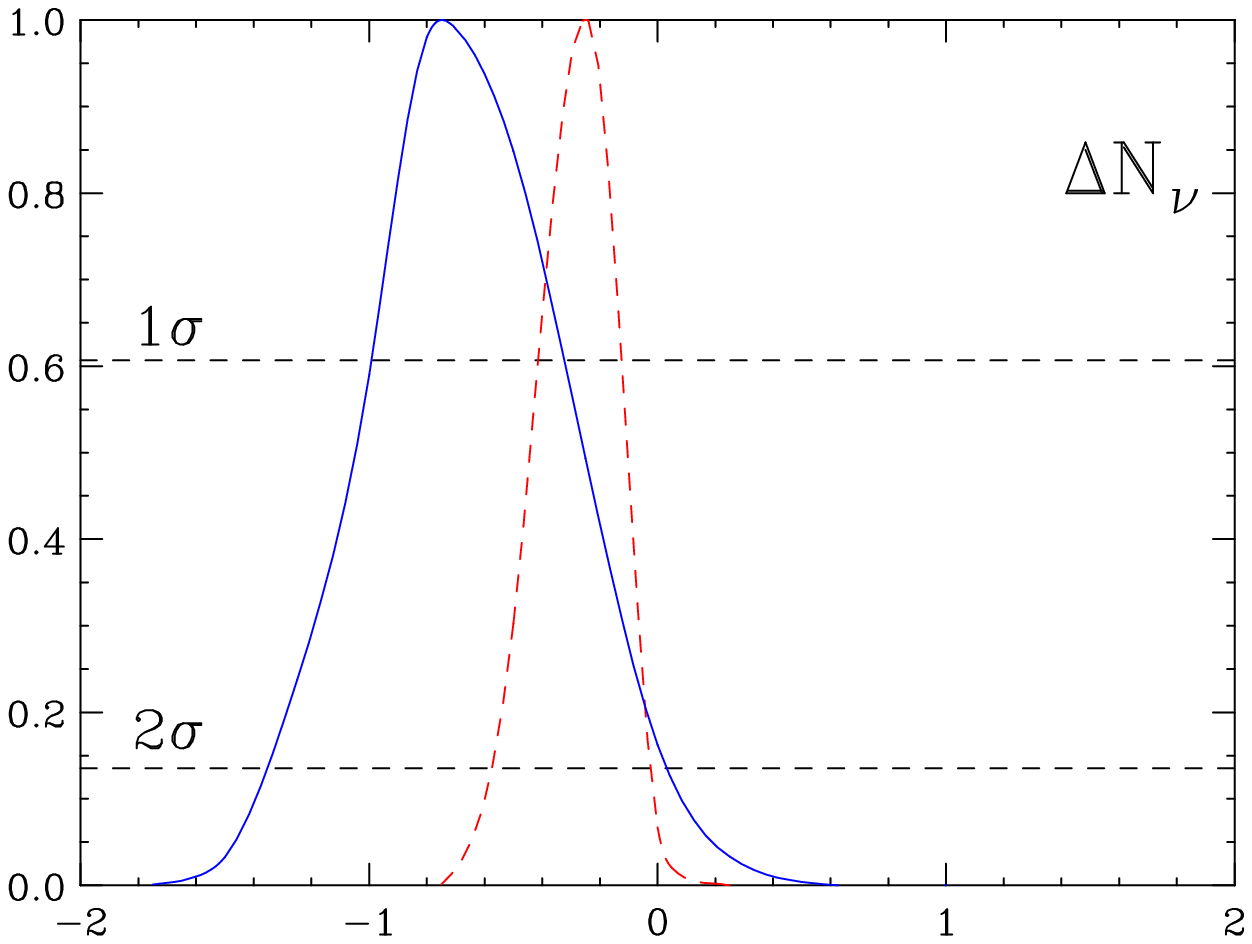} 
      \caption{The marginalized likelihood distributions for \Deln
                from the joint WMAP and BBN analysis 
               for two choices of the primordial abundance of \4he 
               (solid: OSW, dashed: IT). 
      \label{fig:deltannulikelihood}} 
 
\end{center} 
\end{figure} 
 
\begin{figure}[htbp] 
\begin{center} 
\epsfxsize=3.4in 
\epsfbox{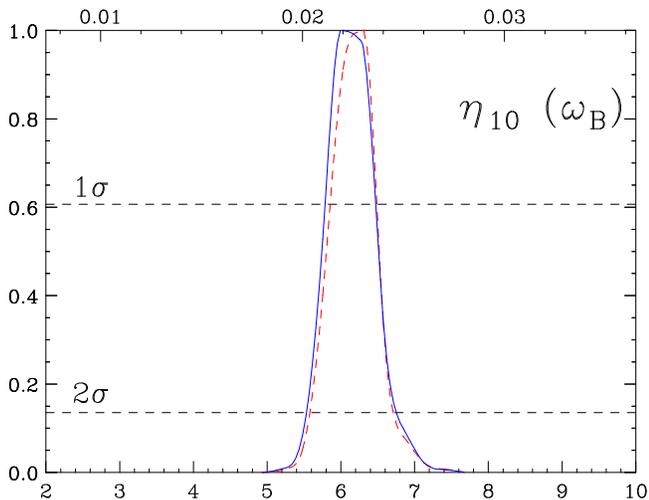} 
      \caption{The marginalized likelihood distributions for $\eta_{10}$ 
                from the joint WMAP and BBN analysis 
               for two choices of the primordial abundance of \4he (solid: 
               OSW, dashed: IT). 
      \label{fig:etalikelihood}} 
 
\end{center} 
\end{figure}

Our results are summarized in Tables I and II and 
Figs.~\ref{fig:deltannulikelihood} and~\ref{fig:etalikelihood}. 
BBN and the primordial D abundance combine to provide a quite accurate 
determination of the baryon density ($0.020 ~\la \omega_{\rm B} ~\la 
0.025$ at $2\sigma$).  The currently large uncertainty in the primordial 
abundance of \4he is responsible for the larger allowed range of $N_{\nu}$. 
While the best fit value for \Nnu is $< 3$ for either the OSW or the IT 
\4he abundances, each are consistent with \Nnu = 3 at $\sim 2\sigma$. 
These results are in excellent agreement with those of Hannestad 
\cite{others}, with which they are most directly related. 
 
Clearly, BBN constrains \Nnu much more stringently than WMAP, while 
the measurement of $\eta_{10}$ by WMAP is at a precision superior 
(by a factor $\sim 2$) to that from BBN.  In this sense, the CBR and 
BBN are quite complimentary.  Indeed, while the constraint on \Nnu 
barely changes with the  inclusion of the WMAP data in a joint analysis 
with BBN, it is sensitive to the adopted \4he abundance;
see Fig.~\ref{fig:deltannulikelihood}. 
On the other hand, the joint constraint 
on $\eta$ is extremely insensitive to the choice of the \4he abundance, 
being dominated by the WMAP data (and the primordial D abundance); see 
Fig.~\ref{fig:etalikelihood}.  
However, BBN and WMAP do provide a very important consistency 
check of the standard model of cosmology at widely separated epochs, 
using significantly different physics.  The excellent agreement between 
\Nnu and $\eta_{10}$ when the universe was 20 minutes and 380,000 years 
old is a major triumph for the (new) standard model of cosmology. 
 
\acknowledgments 
Extensive computations were carried out on the CONDOR system at the 
University of Wisconsin, Madison with parallel processing on up to 
200 CPUs. We thank S. Dasu, W. Smith, D. Bradley and S. Rader for 
providing access to and assistance with CONDOR.  This research was 
supported by the U.S.DOE under Grants No. DE-FG02-95ER40896, 
No. DE-FG02-91ER40676, No. DE-FG02-02ER41216, and No. DE-FG02-91ER40690, 
by the NSF under Grant No. PHY99-07949 and by the Wisconsin Alumni 
Research Foundation.  VB, DM and GS thank the Kavli Institute for 
Theoretical Physics at the University of California, Santa Barbara 
for its support and hospitality.

%\end{references} 
 

\begin{thebibliography}{99} 
%\begin{references} 
 
\bibitem{map} 
C.~L.~Bennett {\it et al.}, 
 astro-ph/0302207. 
%%CITATION = ASTRO-PH 0302207;%% 
 
\bibitem{k} D. Kirkman, D. Tytler, N. Suzuki, J. O'Meara, \& D. 
Lubin, astro-ph/0302006 (2003). 
 
\bibitem{ssg} G. Steigman, D. N. Schramm, \& J. E. Gunn, Phys. Lett. 
{\bf 66B}, 202 (1977). 
 
\bibitem{rs} L. Randall \& R. Sundrum, Phys. Rev. Lett. {\bf 83}, 
3370 (1999); Phys. Rev. Lett. {\bf 83}, 4690 (1999). 
 
\bibitem{rs1} J. M. Cline, C. Grojean, \& G. Servant, Phys. Rev. Lett. 
{\bf 83}, 4245 (1999). 
 
\bibitem{rs2} P. Binetruy, C. Deffayet, U. Ellwanger, \& D. Langlois, 
Phys. Lett. {\bf B477}, 285 (2000). 
 
\bibitem{rs3} R. Maartens, D. Wands, B. Bassett, \& I. Heard, Phys. 
Rev, {\bf D62}, 041301 (2000). 
 
\bibitem{rs4} D. Langlois, R. Maartens, M. Sasaki, \& D. Wands, 
Rev, {\bf D63}, 084009 (2001). 
 
\bibitem{rs5} S. Mizuno \& K. Maeda, Phys. Rev. {\bf D64}, 123521 
(2001). 
 
\bibitem{rs6} J. D. Barrow \& R. Maartens, Phys. Lett. {\bf B532}, 
155 (2002). 
 
\bibitem{rs7} K. Ichiki, M. Yahiro, T. Kajino, M. Orito, \& G. J. 
Mathews, Phys. Rev. {\bf D66}, 043521 (2002). 
 
\bibitem{bratt} J. D. Bratt, A. C. Gault, R. J. Scherrer, \& T. P. 
Walker, Phys Lett {\bf B546}, 19 (2002) 
 
\bibitem{hu} 
For a recent review see, 
W.~Hu and S.~Dodelson, 
 astro-ph/0110414. 
%%CITATION = ASTRO-PH 0110414;%% 
 
\bibitem{camb} 
A.~Lewis, A.~Challinor and A.~Lasenby, 
Astrophys.\ J.\  {\bf 538}, 473 (2000) 
[astro-ph/9911177]; 
%%CITATION = ASTRO-PH 9911177;%% 
http://camb.info/ 
 
\bibitem{cmbfast} 
U.~Seljak and M.~Zaldarriaga, 
Astrophys.\ J.\  {\bf 469}, 437 (1996) 
[astro-ph/9603033]. 
%%CITATION = ASTRO-PH 9603033;%% 
 
\bibitem{inflation} 
A.~H.~Guth, Phys.\ Rev.\ D {\bf 23}, 347 (1981). 
%%CITATION = PHRVA,D23,347;%% 
 
\bibitem{HST} 
W.~L.~Freedman {\it et al.}, 
Astrophys.\ J.\  {\bf 553}, 47 (2001) 
[astro-ph/0012376]. 
%%CITATION = ASTRO-PH 0012376;%% 
 
\bibitem{chaboyer} 
L.~M.~Krauss and B.~Chaboyer, Science {\bf 299}, 65 (2003). 
%%CITATION = ASTRO-PH 0111597;%% 
 
\bibitem{maptt} 
G.~Hinshaw {\it et al.}, 
 astro-ph/0302217. 
%%CITATION = ASTRO-PH 0302217;%% 
 
\bibitem{mapte} 
A.~Kogut {\it et al.}, 
 astro-ph/0302213. 
%%CITATION = ASTRO-PH 0302213;%% 
 
\bibitem{mapcode} 
L.~Verde {\it et al.}, 
 astro-ph/0302218. 
%%CITATION = ASTRO-PH 0302218;%% 
 
\bibitem{others} 
P.~Crotty, J.~Lesgourgues and S.~Pastor, 
astro-ph/0302337; 
%%CITATION = ASTRO-PH 0302337;%% 
E.~Pierpaoli, 
astro-ph/0302465; 
%%CITATION = ASTRO-PH 0302465;%% 
S.~Hannestad, 
astro-ph/0303076. 
%%CITATION = ASTRO-PH 0303076;%% 
 
\bibitem{kssw} see, \eg 
J. P. Kneller, R. J. Scherrer, G. Steigman, \& T. P. 
Walker, Phys. Rev. {\bf D64}, 123506 (2001). 
 
\bibitem{sperg} D. N. Spergel, \etal astro-ph/0302209. 
 
\bibitem{knellsteig} J. P. Kneller \& G. Steigman, Phys. Rev. {\bf D67}, 
063501 (2003). 
 
\bibitem{ks} H. S. Kang \& G. Steigman, Nucl. Phys. {\bf B372}, 
494 (1992). 
 
\bibitem{els} R. Epstein, J. Lattimer, \& D. N. Schramm, Nature {\bf 263}, 
198 (1976). 
 
\bibitem{bta} S. Burles \& D. Tytler, ApJ {\bf 499}, 699 (1998a). 
 
\bibitem{btb} S. Burles \& D. Tytler, ApJ {\bf 507}, 732 (1998b). 
 
\bibitem{o'm} J. M. O'Meara, \etal, ApJ {\bf 552}, 718 (2001). 
 
\bibitem{pb} M. Pettini \& D. V. Bowen, ApJ {\bf 560}, 41 (2001). 
 
\bibitem{dod} S. D'Odorico, M. Dessauges-Zavadsky, \& P. Molaro, 
A\&A {\bf 338}, L1 (2001). 
 
\bibitem{gs01} G. Steigman, To appear in the Proceedings of the 
STScI Symposium, "The Dark Universe: Matter, Energy, and Gravity" 
(April 2 -- 5, 2001), ed. M. Livio [astro-ph/0107222]. 
 
\bibitem{lev01} S. A. Levshakov, M. Dessauges-Zavadsky, S. D'Odorico, 
\& P. Molaro, ApJ {\bf 565} 696 (2002). 
 
\bibitem{lev1} S. A. Levshakov, W. H. Kegel \& F. Takahara, ApJ 
{\bf 499}, L1 (1998). 
 
\bibitem{lev2} S. A. Levshakov, W. H. Kegel \& F. Takahara, A\&A 
{\bf 336}, 29L (1998). 
 
\bibitem{lev3} S. A. Levshakov, W. H. Kegel \& F. Takahara, MNRAS 
{\bf 302}, 707 (1999). 
 
\bibitem{itl} Y. I. Izotov, T. X. Thuan, \& V. A. Lipovetsky, ApJS 
{\bf 108}, 1 (1997). 
 
\bibitem{it} Y. I. Izotov \& T. X. Thuan, ApJ {\bf 500}, 188 (1998). 
 
\bibitem{os} K. A. Olive \& G. Steigman, ApJS {\bf 97}, 49 (1995). 
 
\bibitem{oss} K. A. Olive, E. Skillman, \&  G. Steigman, ApJ {\bf 483}, 
788 (1997). 
 
\bibitem{fo} B.D. Fields \& K. A. Olive, ApJ {\bf 506}, 177 (1998). 
 
\bibitem{ppr} M. Peimbert, A. Peimbert, \& M. T. Ruiz, ApJ {\bf 541}, 
688 (2000). 
 
\bibitem{gsv} R. Gruenwald, G. Steigman, \& S. M. Viegas, ApJ {\bf 567}, 
931 (2002). 
 
\bibitem{vgs} S. M. Viegas, R. Gruenwald, \& G. Steigman, ApJ {\bf 531}, 
813 (2000). 
 
\bibitem{sj} D. Sauer \& K. Jedamzik, A\&A {\bf 381}, 361 (2002). 
 
\bibitem{osw} K. A. Olive, G. Steigman, \& T. P. Walker, Phys. Rep. 
{\bf 333}, 389 (2000). 
 
\bibitem{lsnd} 
A.~Aguilar {\it et al.}  [LSND Collaboration], 
Phys.\ Rev.\ D {\bf 64}, 112007 (2001) 
[hep-ex/0104049]. 
%%CITATION = HEP-EX 0104049;%% 
 
\bibitem{paul} 
P.~Langacker, UPR-0401T. 
 
\bibitem{pdg} 
K.~Hagiwara {\it et al.}  [Particle Data Group Collaboration], 
Phys.\ Rev.\ D {\bf 66}, 010001 (2002). 
%%CITATION = PHRVA,D66,010001;%% 
 
\bibitem{css} X. Chen, R. J. Scherrer, \& G. Steigman, Phys. Rev. 
{\bf D63}, 123504 (2001). 
 
\end{thebibliography}
\end{document}